\documentclass[prl,aps,amsfonts,amssymb,twocolumn,floats,showpacs]{revtex4}
\usepackage{epsf}
\usepackage{graphicx}

\begin{document}
\title{Multifractal zero mode for disordered graphene}

\author{I. Kleftogiannis and S.N. Evangelou\footnote{e-mail:sevagel@cc.uoi.gr}}
\affiliation{Department of Physics, University of Ioannina,
Ioannina 45110, Greece}

\begin{abstract}
Off-diagonal disorder with random hopping between the sublattices of a bipartite lattice is described by a Hamiltonian which has chiral (sub-lattice) symmetry. The energy spectrum is symmetric around $E=0$ and for odd total number of lattice sites an isolated zero mode always exists, which coincides with the mobility edge of an Anderson transition in two dimensions($2D$). In the chiral orthogonal symmetry class BDI we compute the fractal dimension $D_{2}$ of the zero mode for graphene samples with edges. In the absence of disorder $D_{2}=1$, which corresponds to a one-dimensional edge states, while for strong disorder $D_{2}$ decays towards $0$ and the zero mode becomes localized. The  similarities and differences between zero modes in the honeycomb and the square bipartite lattices are pointed out. 
\end{abstract}


\maketitle

\section{I. Introduction}

\par
\medskip
The interest in zero energy modes extends to more than a decade and to all areas of physics\cite{r1}. It was recently revived owing to the emergence of graphene\cite{r2} and topological insulators\cite{r3}. The zero modes are believed to signal non-trivial lattice topology and are connected to excitations of fractional charge\cite{r1}. They also exist in the spectrum of a bipartite lattice (two interconnected sublattices) if the total number of lattice sites is odd\cite{r4}. In the presence of off-diagonal disorder which connects the two sublattices, in the so-called chiral disordered systems, a zero mode is a critical point (mobility edge) of the Anderson metal-insulator transition. The Dirac point of graphene is a stable zero mode in two dimensions ($2D$) which becomes a critical point for appropriate off-diagonal disorder, and zero modes with half-integer charge appear in superconducting Bogolyubov-de Gennes systems as Majorana fermions\cite{r5}.  

\par
\medskip
Graphene  has attracted immense interest in the last decade. In this novel material the electrons can be described by Dirac equation and its band structure displays two valleys (act as pseudo spin states) related by time-reversal\cite{r2}. The Fermi surface has two non-equivalent Dirac points so graphene is a semimetal or a zero gap semiconductor, at long length scales it has a linear energy dispersion with a density of states which goes to zero at $E=0$\cite{r2}. On the other hand, disorder causes Anderson localization for all states in $2D$ disordered systems, at least in the  limit of infinite size\cite{r6}. This is expected also for graphene since the honeycomb lattice has a coordination number even smaller than the square lattice.

\par
\medskip
Our interest here is the so-called off-diagonal disorder, which appears in the random hopping among the sites of two interconnected sublattices $A$, $B$ (mirror images of one another), of a bipartite $2D$ lattice such as the honeycomb lattice of graphene. For graphene off-diagonal disorder between nearest neighbours (n.n.) can appear  by removing lattice bonds at random, it can simulate ripples, etc. It is expected (as diagonal disorder) to mix graphene's two-valleys and if correlated to break time-reversal symmetry by decoupling valleys, etc. The off-diagonal disorder (random hopping between $A$ and $B$) is different from usual diagonal disorder (random site energies), in the sense that it respects the discrete chiral (sublattice) symmetry\cite{r7}. The chiral disordered systems were classified by three random matrix ensembles (the chiral universality classes BDI, AIII, CII)\cite{r8}. A zero mode can exist at $E=0$\cite{r9} where the chiral system undergoes a metal-insulator transition in $2D$ driven by topologically induced Anderson localization\cite{r10}. This type of localization is shown via non-perturbative field-theoretic techniques and is controlled by vortex-like excitations of the sigma model, via a mechanism similar to the Berezinskii-Kosterlitz-Thouless transition\cite{r11}.  

\par
\medskip
The chiral symmetry of the Hamiltonian $H$, with random hopping among the $A$, $B$ sublattices, manifests itself in the energy spectrum which is composed of random energy pairs $(E_{j},-E_{j})$, where $j$ labels the eigenvalues of the same sign. For an odd total number of lattice sites the spectrum is symmetric  around $E=0$ and allows zero modes which cannot match, as required by chiral symmetry, any of the positive or negative $E_{j}$'s. In fact, the $E=0$ mode  matches to itself. A zero mode is a critical state which emerges among localized states for topological reasons, the localization length (and the density of states) diverges at $E=0$\cite{r12,r13}. In $2D$ off-diagonal disorder can localize all non-zero states, the only exception is the zero mode\cite{r14}. We shall show that off-diagonal disorder can localize even the zero mode when it becomes strong\cite{r10}. In the original treatment of chiral systems\cite{r12} no traces of weak localization (quantum interference) was found to all orders in perturbation theory. We show that the critical zero mode displays multifractal fluctuations\cite{r15} described by a fractal dimension $D_{2}$, which tends to zero as the disorder becomes very large. In other words, for strong off-diagonal disorder the zero mode state becomes localized in agreement with \cite{r10}. 
 
\par
\medskip 
Our purpose is to study the multifractal properties of the zero mode in the absence of magnetic field or spin-orbit coupling, that is for a Hamiltonian $H$ with real random hopping among $A$ and $B$ which belongs to the chiral orthogonal class $BDI$. This $H$ has a mobility edge in $2D$ which is a zero mode. In the absence of chiral symmetry in $2D$ a mobility edge  can exist only in two circumstances, if the usual orthogonal universality class is replaced by the symplectic with spin-orbit coupling, or by the unitary class with added magnetic field. In the presence of disorder two mobility edges usually appear as critical points of an Anderson transition. For a chiral $2D$ system they merge into one at the zero mode whose origin is topological\cite{r9}. Our study has an advantage over $3D$, where mobility  edges usually exist, since they are more difficult to handle. For the chiral class BDI in $2D$ we compute the fractal dimension $D_{2}$\cite{r16} of the zero mode state. This is done in order to see the fate of the Dirac point of graphene in the presence of off-diagonal disorder. One might also ask the question about the difference of the obtained $D_{2}$ between a honeycomb and a square lattice, since both lattices with appropriate off-diagonal disorder belong to the same chiral class BDI. Another question addressed in this paper is whether the critical zero mode localizes for strong disorder and the nature of its localization. 


\section{II. The $2D$ Hamiltonian with off-diagonal disorder}

\par
\medskip
The honeycomb lattice consists of two interconnected triangular sublattices $A$ and $B$. If the lattice has zigzag(zz) edges zero modes can appear located at the zz boundary of one sublattice\cite{r17}. A spinless electron with off-diagonal disorder is described by the single particle nearest neighbour tight-binding Hamiltonian (belongs to the chiral orthogonal class BDI): 
\begin{equation}
      H=-\sum_{{r}\in A}\sum_{i=1,2,3}t_{{r},i}\left (a_{{r}}^{\dag}b_{{r}+{s}_{i}}+b_{{r}+{s}_{i}}^{\dag}a_{{r}}\right),
\end{equation}
$a_{{r}}$, $b_{{r}+{s}_{i}}^{\dag}$ annihilates(creates) an electron in $A$ or $B$ sublattice. The $A$, $B$ sublattices are interconnected by random hopping $t_{{r},i}$,$\;{r}\in A$ sublattice and $s_{i}, i=1,2,3$ connects to its three nearest neighbours (n.n.) which belong to the $B$ sublattice. The nearest neighbour hopping $t_{{r},i}$ are chosen real uncorrelated random variables distributed by the law:
\begin{equation}
P(t)=\frac{1}{Wt}, \; \; t\in[e^{-\frac{W}{2}}, e^{+\frac{W}{2}}],
\end{equation}
which guarantees  positive $t$ only (its mean $<t>\neq 0$). This is equivalent to a box distribution in the range $[-W/2,+W/2]$ for the logarithm $\ln(t)$. Our choice of n.n. off-diagonal disorder (its strength is denoted by $W$) respects the chiral (sublattice) symmetry, while for pure graphene $t_{{r},i}=t$ for every $r$, $i$ ($W=0$). A local symmetric gauge model of off-diagonal disorder (also chiral) has zero mean $<t>=0$, $t$ not $\ln(t)$ is chosen, e.g. from a box distribution of width $W$, which gives a single value of off-diagonal disorder (no other energy scale exists in the Hamiltonian and for the mean zero hopping the lattice vanishes). In the gauge model all ensemble averages are invariant under sign changes of the wave function and the single particle averages are site-diagonal\cite{r12}. The Hamiltonian of Eq.(1) for both models of n.n. off-diagonal disorder (logarithmic and zero mean) has no spin-dependence, it is invariant under time-reversal ($t$ is real), and it respects chiral symmetry (belongs to BDI).

\par
\medskip
A bipartite lattice consists of the two sublattices $A$ and $B$, in the $A$-$B$ basis the Hamiltonian is:
\begin{equation}
H=\left[\begin{array}{cc}
0 & H_{AB}\\
H_{AB}^{+} & 0\end{array}\right],
\end{equation}
$H_{AB}$ contains the random hopping between
$A$ and $B$. The chiral symmetry can be expressed in
the anti-commutation relation $\left[H,\sigma_{3}\right]=0$ with
the pseudospin matrix $\sigma_{3}= \left[\begin{array}{cc}
1 & 0\\
0 & -1\end{array}\right]$.
For an energy $E$ and corresponding state 
$\Psi=\left(\begin{array}{c}
\Psi_{A}\\\Psi_{B}\end{array}\right)$, $H\Psi=E\Psi$, via the above commutation relation an energy $-E$ and a state $\sigma_{3}\Psi=\left(\begin{array}{c}
\Psi_{A}\\-\Psi_{B}\end{array}\right)$ also exists, $H(\sigma_{3}\Psi)=-\sigma_{3}H\Psi=-E(\sigma_{3}\Psi)$. 
The presence of chiral symmetry implies pairs of energies $E,-E$
with wave functions $\Psi,\sigma_{3}\Psi$, respectively.
The spectrum is symmetric spectrum around $E=0$ and for an odd number of lattice sites 
a state with zero energy 
\begin{equation}
\left(\begin{array}{c}
\Psi_{A}\\
\Psi_{B}\end{array}\right)
=\left(\begin{array}{c}
\Psi_{A}\\
-\Psi_{B}\end{array}\right)
\Rightarrow \begin{array}{c}
\Psi_{A}\neq 0\\
\Psi_{B}=0\end{array}.
\end{equation}
and its amplitude has support only on the $A$ sublattice being zero on the $B$ sublattice. 

\par
\medskip
The Hamiltonian $H$ has at least $N_{A}-N_{B}$ independent zero modes\cite{r7}, $N_{A}$ is the number of sites for sublattice $A$ and $N_{B}$ the number of sites for sublattice $B$, with $N_{A}>N_{B}$.
The total number of lattice sites $N=N_{A}+N_{B}$ is odd and since the number
of energies is also odd the chiral symmetry leads to an $E=0$ mode. We choose $N_{A}=N_{B}+1$ for one zero mode
to exist, which can be easily obtained\cite{r4,r16} by solving:
\begin{equation}
H\Psi=0\Leftrightarrow\left[\begin{array}{cc}
0 & H_{AB}\\
H_{AB}^{+} & 0\end{array}\right]\left(\begin{array}{c}
\Psi_{A}\\
\Psi_{B}\end{array}\right)=0\Leftrightarrow\left\{ \begin{array}{c}
H_{AB}\Psi_{B}=0\\
H_{AB}^{+}\Psi_{A}=0\end{array}\right.\end{equation}
The  $\Psi_{A}$ are obtained from the homogeneous system $H_{AB}^{+}\Psi_{A}=0$
of $N_{B}$ linear  equations defined on the $B$ sublattice, for the amplitudes on the $A$ sublattice. The system has $N_{A}=N_{B}+1$ unknowns (the additional equation is provided by the normalization)\cite{r15}. 

\par
\medskip
The critical zero mode state of Eq.(1) is multifractal, it has infinite self-similar detail which roughly repeats itself. In the renormalization group sense multifractality implies the presence of infinitely many relevant operators. The critical wave functions at the disorder induced Anderson transition are multifractals\cite{r15}. For weak disorder the periodic Bloch states become chaotic and for strong disorder (above a critical disorder for  dimensions higher than two) become localized in a region of the order of the localization length $\xi$. A finite $\xi$ for a localized state can be estimated by the participation ratio (outside the localization region denoted by $\xi$ the states usually decay exponentially). At  the zero mode the localization length diverges logarithmically and precisely at $E=0$ takes any value which depends strongly on the choice of boundaries\cite{r4}. At the Anderson transition a critical wave function lies at the mobility edge which separates extended (chaotic) from localized states. The multifractal states in the border between chaotic and localized states have filamentary structure and are described by multifractal dimensions\cite{r15}. In the chiral problem studied, however, no chaotic (diffusive) states exist and the critical zero mode emerges among localized $2D$ states only. 

\par
\medskip
The multifractals are inhomogeneous fractals described by many (a whole spectrum) of fractal dimensions $D_{q}, q\in\left[-\infty,\infty\right]$, obtained from scaling the distributions for a physical quantity rather than its mean (ordinary fractals are characterized by a single fractal dimension $D$ via scaling of the mean). Alternatively, the multifractals describe a singularity spectrum $f(\alpha)$, it is related to $D_{q}'s$ by an appropriate Legendre transform. In our case the irregularly distributed probability amplitude $|\psi_{l,m}|^{2}$ at the lattice site $l,m$ of the normalized  ($\sum_{l,m} |\psi_{l,m}|^{2}=1$)  zero mode state is collected independently of $l,m$. In the presence of chiral symmetry a state with energy $E$: $\psi_{l,m}$, $l,m=1,2,...$ along the sites of the bipartite lattice, at the mirror image energy $-E$ corresponds to the state:  $(-1)^{l+m}\psi_{l,m}$. The  $|\psi_{l,m}|^{2}$ is not self-averaging and can be described by the spectrum of dimensions obtained from scaling of all the moments of the wave function. In order to describe $|\psi_{l,m}|^{2}$ of the zero mode we introduce multifractal measures (the moments of the wave function), in other words one must compute the spectrum of dimensions
\begin{equation}
D_{q}={\frac {1}{1-q}}\lim_{L\rightarrow\infty} {\frac{\ln\sum_{l,m}|\psi_{l,m}|^{2q}}{\ln L}}, 
\;\; q\in\left[-\infty,\infty\right],
\end{equation}
with respect to the linear length $L$ of the square sample. In the rest we focus only on the fractal dimension $D_{2}$ for $q=2$. The $D_{2}$ (known as correlation dimension) describes scaling of the inverse participation ratio 
\begin{equation}
IPR=\sum_{l,m}\bigl|\psi_{l,m}\bigr|^{4}\sim L^{-D_{2}},
\end{equation}
for large $L$. The $IPR$ gives information for the degree of localization (it is the inverse for the number of lattice sites participating in the state). The $D_{2}$ is conveniently obtained by scaling the averaged $<IPR>$, or the typical $\exp <\ln(IPR)>$, {\it vs.} the linear size $L$, $D_{2}$ is the slope in the corresponding log-log plot. For a metal a chaotic (extended) state has $D_{2}=d$, the space dimension, for a localized state $D_{2}\to 0$ (insulator), while for a multifractal critical state (at the metal-insulator transition) $D_{2}$ lies in between these limits. 
The multifractal dimensions $D_{q}$ for the wave function are closely related to scaling for the moments of the local density of states, e.g. for the zero mode they can represent correlation functions for the local density of states.

\section{III. The zero mode for the honeycomb lattice}

\par
\medskip
We have solved the linear equations of Eq.(5) for the honeycomb lattice which consists of $N_{c}$ horizontal zig-zag chains (the width of the lattice is $\frac{N_{c}}{2}\sqrt{3}$ in units of $a=\sqrt{3}a_{C-C}$,  $a_{C-C}$ is the distance between carbon atoms) and $N_{sc}$ the number of sites for each chain (the lattice length is $L=\frac{N_{sc}-1}{2}$ in units of $a$). The lattice has $N=N_{c} N_{sc}$ sites, $N$ is odd if $N_{c}$ and $N_{sc}$ are odd. We have also taken $N_{c}\sqrt{3}\sim N_{sc}-1$ which corresponds to almost square $L\times L$ graphene samples without boundary conditions (bc). The chosen samples have zz edges in the two parallel boundaries and in the other two have armchair edges. The amplitudes of the zero mode wave function  are obtained by building a statistical ensemble of random Hamiltonians for every value of the off-diagonal disorder $W$. For $W=0$ the lattice symmetry with zz edges guarantees the presence of zero mode edge states\cite{r17}. These zero mode states appear due to the lattice topology (the Hilbert space structure), for diagonal disorder see \cite{r18}. Thus, the zero mode is an edge state having amplitude at the zz boundary and fractal dimension $D_{2}=1$. For non-zero off-diagonal disorder $W$ which respects the chiral symmetry the fractal dimension $D_{2}$ diminishes. For strong $W$ the zero mode state localizes along the zz edges. 

\begin{figure}[htb]
\centering
\rotatebox{-90}{\includegraphics[scale=0.33]{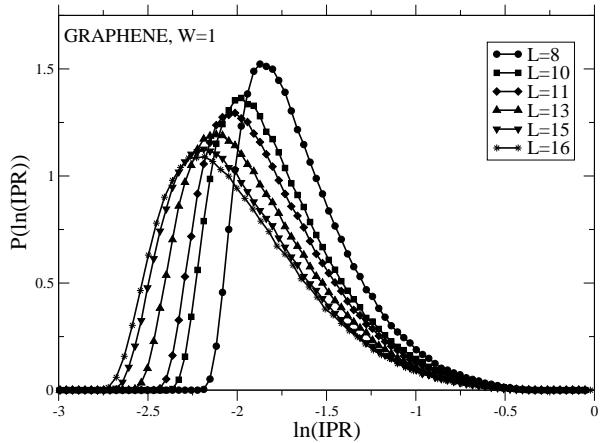}}
\caption{The flow of the probability distribution $P(\ln(IPR))$ obtained from half a million graphene squares of linear lengths about $L=8,10,11,13,15,16$ with logarithmic off-diagonal disorder of strength $W=1$.}
\label{Fig.1}
\end{figure}

\par
\medskip
In Fig.1 we plot the distribution of the $IPR$ for the zero mode state.  In Fig.2 
the mean $<IPR>$ and the geometric mean $\exp(<\ln(IPR)>)$ are plotted in a log-log plot
$vs. $ the linear size $L$. From the slopes one can obtain the fractal dimension $D_{2}$, for zero disorder $D_{2}=1$ 
and for strong disorder (large $W$) it decreases to zero ($D_{2}\to 0$). For graphene the localization of the zero mode wave function occurs on the edge. For finite values of $W\neq 0$ the fractal dimension $D_{2}$ takes values between zero and one.

\begin{figure}[htb]
\centering
\rotatebox{-90}{\includegraphics[scale=0.33]{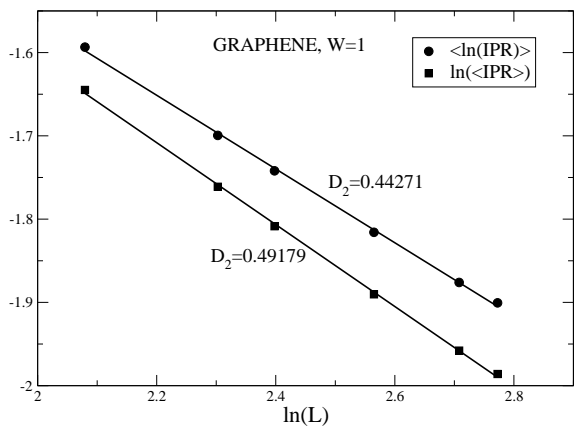}}
\caption{The log-log plot for the scaling of the mean $<IPR>$ and the typical $\exp<\ln(IPR)>$ with graphene samples of linear size $L$. The data are taken for half a million realizations with logarithmic off-diagonal disorder $W=1$, and the straight lines are the fits which give the fractal dimensions $D_{2}$.}
\label{Fig2}
\end{figure}

\section{IV. Comparison with the square lattice}

\par
\medskip
The above analysis applies to any bipartite lattice (also for the square) with appropriate off-diagonal disorder which respects chiral symmetry. This is true as long as appropriate boundary conditions are taken and the size allows presence of the zero mode\cite{r9}, e.g. vanishes by taking periodic boundary conditions. We repeated our computations for square lattice $N=L\times L$ samples, with $N$ odd for the $E=0$ mode to appear. In Fig.3 the probability distribution for the $IPR$ is shown. In Fig.4 the log-log plots of the mean and the typical from the slopes  allow to compute the fractal dimension $D_{2}$ for the square lattice. 

\begin{figure}[htb]
\centering
\rotatebox{-90}{\includegraphics[scale=0.33]{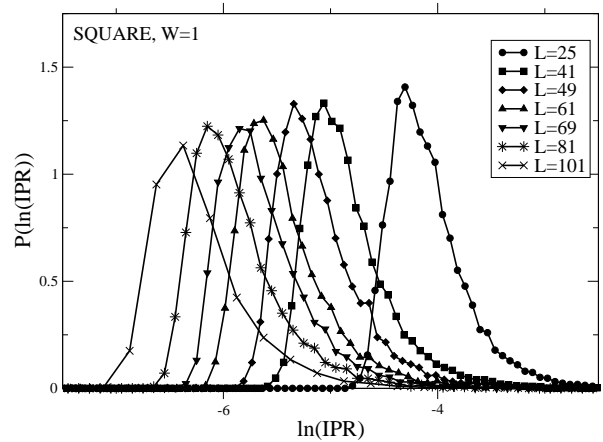}}
\caption{The flow of the probability distribution for the $\ln(IPR)$, taken from half a million realisations of square lattices of lengths $L=25,41,49,61,69,81,101$, with off-diagonal disorder $W=1$.}
\label{Fig.3}
\end{figure}

\par
\medskip
In Fig.5 are shown the results of $D_{2}$ for both bipartite lattices, graphene and square. 
In the absence of disorder ($W=0$) for graphene the zz edges  contribute to a zero edge mode with dimension $D_{2}=1$ while for the square $D_{2}=2$. For low disorder, e.g. $W=0.1$, $D_{2}$ is close to one for graphene and two for the square, respectively. For higher disorder $W$ the $D_{2}$ take values between one and zero for graphene and two to zero for the square. For strong off-diagonal disorder the wave function  becomes localised along a one-dimensional path, e.g on the zz edges for graphene samples, and localized in the bulk for the square\cite{r16}. For strong disorder ($W=5$) the fractal dimension $D_{2}$ approaches zero in both cases. 

\begin{figure}[htb]
\centering
\rotatebox{-90}{\includegraphics[scale=0.33]{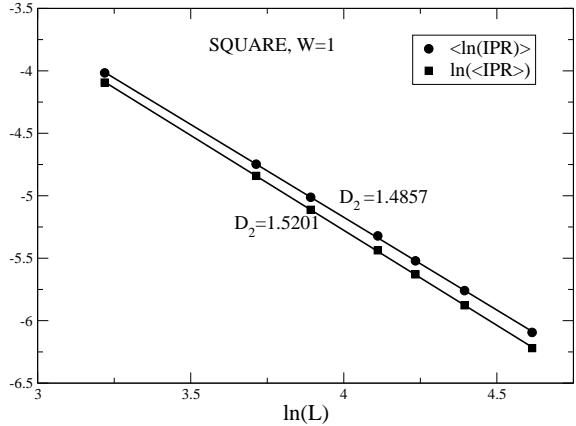}}
\caption{The log of the mean $<IPR>$ and the log of the typical $\exp(<\ln(IPR)>)$, {\it vs.} the log of the linear size $L$. The off-diagonal disorder is $W=1$ for half a million realizations. The straight lines are the fits which give the fractal dimensions $D_{2}$.}
\label{Fig.4}
\end{figure}

\par
\medskip
Our conclusions concern $D_{2}$ for graphene and the square lattice in the presence of off-diagonal disorder both with chiral symmetry. The $D_{2}$ obtained from scaling the average $IPR$ is expected to be more sensitive to rare events than scaling the typical value which turns out to be more reliable in this case. The two $D_{2}$'s, derived form scaling the mean and the typical, are comparable with each other with rather small uncertainty (Fig.5). A rather large difference is seen for disorder higher than $W\simeq 1$. One could denote this disorder as a point of a freezing transition (difference between averaged and typical values)\cite{r19}. The main difference between graphene and square occurs for zero disorder, where for graphene the zero mode state  has a fractal dimension $D_{2}=1$, which corresponds to a wave function localised along the zz edge (edge state), while for the square $D_{2}=2$ for an extended state. For finite values of disorder the $D_{2}$'s in graphene take values between one (the wave function is localised on the edge of the sample) and zero. For strong disorder both systems have $D_{2}\to 0$.

\begin{figure}[htb]
\centering
\rotatebox{-90}{\includegraphics[scale=0.33]{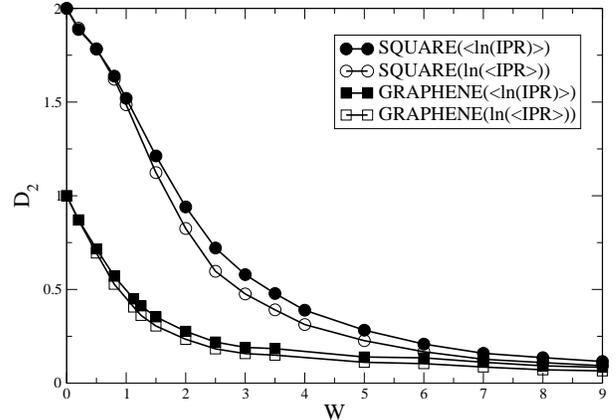}}
\caption{The fractal dimension $D_{2}$ for graphene and the square lattice as a function of off-diagonal disorder $W$. The $D_{2}$ is obtained from the mean $<IPR>$ and the typical $\exp<\ln(IPR)>)$ for each size $L$  (by scaling $\ln<IPR>$ and $<\ln(IPR)>$ {\it vs.} $\ln L$). For $W=0$ the fractal dimension of graphene is $D_{2}=1$ and for the squared lattice is $D_{2}=2$. For strong off-diagonal disorder(large $W$) both $D_{2}$ become zero (the zero mode state localizes).}
\label{Fig.5} 
\end{figure}

\section{V. Discussion-Conclusions}

\par
\medskip
In this paper we have studied the multifractal properties of the zero modes which appear in chiral orthogonal systems (class BDI) for an odd number of lattice sites. The zero mode is simply a zero energy state of the real symmetric Hamiltonian with off-diagonal disorder which respects chiral symmetry and occurs exactly at the critical energy (mobility edge) of an Anderson metal-insulator transition in $2D$ chiral disordered systems. This transition has only insulating and critical ($E=0$) phase, it has no metallic phase. The fractal dimension $D_{2}$ for the critical state at $E=0$ is computed for graphene and the square, from scaling of the inverse participation ratio $IPR$ (averaged $<IPR>$ and typical $\exp <\ln(IPR)>$) {\it vs.} the linear system size $L$. The multifractal zero mode is influenced by the system boundaries even for infinite system size, e.g. for graphene the boundaries are responsible for the edge state structure of the Dirac point in the presence of off-diagonal disorder\cite{r17}. 

\par
\medskip
A chiral system allows the presence of a mobility edge in $2D$ while the scaling theory of localization\cite{r6} forbids critical states in $2D$ orthogonal non-chiral systems. Off-diagonal disorder which respects chiral symmetry allows a topological critical state which is a zero mode in $2D$. The problem of zero modes, apart from its intrinsic interest, is significant in other areas of physics. The first examples were found in the eighties and it is progressively realised that they are important in disordered systems, e.g. they can describe the low-energy physics of disordered spin chains, which can be mapped into disordered fermions by Jordan-Wigner transformation. Our findings support the idea that the origin of the zero modes is geometric. Ther multifractality is shown in two chiral disordered $2D$ systems, for the honeycomb and the square lattice. The obtained zero modes are found to behave rather similarly, unless finite size effects are present. Their difference depends strongly on the topology of the perimeter, e.g in graphene for no disorder the zz edges on the perimeter can lead to zero energy modes. 

\par
\medskip
In conclusion, for an odd number of lattice sites a topological zero mode exists for chiral disordered systems which is a mobility edge of an Anderson transition in $2D$. We have compared its multifractal properties for graphene and the square lattice with n.n. off-diagonal disorder. In both cases the isolated critical zero mode emerges within the localized $2D$ states and has amplitude only on one of the interconnected sublattices $A$ or $B$. The complex quantum interference in disordered systems which usually leads to Anderson localization is absent in chiral systems. In ref. \cite{r12} a perturbation theory found no traces of localization for chiral systems, however, non-perturbative effects\cite{r10} showed topological localization. We confirm these results. The zero mode has no topological protection against localization which occurs for strong off-diagonal disorder ($D_{2}\to 0$). Our conclusions should remain valid for other chiral systems, e.g. if the time-reversal invariance is broken by a magnetic field and the invariance over spin-rotation by spin-orbit coupling. An old idea, whether physics can be reduced to geometry, is supported by the presence of lattice topology and multifractality in disordered systems. In $2D$ topological phenomena (non-perturbative) exist, e.g. the quantum Hall effect\cite{r15}, etc. We have shown Anderson localization in chiral disordered systems appears in a topological sense, e.g. at the zz edges of graphene. The presence of zero modes is topologically protected by chiral symmetry but  their criticality is not protected since they localise for strong disorder. The zero modes are also connected to fractional charges and monopoles\cite{r20}. The obvious extensions of this work is to obtain the whole spectrum of multifractal dimensions $D_{q}, \;q\in\left[-\infty,\infty\right]$, which they are expected to be a non-trivial function of $q$, and generalize this study to the other chiral universality classes, AIII, CII\cite{r21,r22}.

\end{document}